\documentclass[twocolumn,floatfix,superscriptaddress]{revtex4-2}
\usepackage{amsmath}
\usepackage{graphicx}
\usepackage{epsfig}
\usepackage{color}
\usepackage{hyperref}
\hypersetup{colorlinks=true, citecolor=blue, urlcolor=blue, linkcolor=blue}
\usepackage{cancel}
\usepackage[normalem]{ulem}
\usepackage[utf8]{inputenc}
\usepackage{amssymb}
\usepackage{hyphenat}

\begin{document}

\title{Finite-size correlation behavior near a critical point: a simple metric for monitoring the state of a neural network}
 
\author{Eyisto J. Aguilar Trejo}
\affiliation{Instituto de Ciencias F\'isicas (ICIFI-CONICET), Center for Complex Systems and Brain Sciences (CEMSC3), Escuela de Ciencia y Tecnología, Universidad Nacional de Gral. San Martín, Campus Miguelete, 25 de Mayo y Francia,  1650, San Martín, Buenos Aires, Argentina}

\affiliation{Consejo Nacional de Investigaciones Cient\'ificas y T\'ecnicas (CONICET), Godoy Cruz 2290, 1425,  Buenos Aires, Argentina}

\author{Daniel A. Martin}
\email{dmartin@unsam.edu.ar}
\affiliation{Instituto de Ciencias F\'isicas (ICIFI-CONICET), Center for Complex Systems and Brain Sciences (CEMSC3), Escuela de Ciencia y Tecnología, Universidad Nacional de Gral. San Martín, Campus Miguelete, 25 de Mayo y Francia,  1650, San Martín, Buenos Aires, Argentina}

\affiliation{Consejo Nacional de Investigaciones Cient\'ificas y T\'ecnicas (CONICET), Godoy Cruz 2290, 1425,  Buenos Aires, Argentina}

\author{Dulara De Zoysa}
\affiliation{College of Computer, Mathematical, and  Natural  Sciences, University of  Maryland, College Park, MD, USA}

\author{Zac Bowen}
\affiliation{Fraunhofer USA Center Mid-Atlantic, Riverdale, MD 20737, USA}

\author{Tomas S. Grigera} 
\affiliation{Consejo Nacional de Investigaciones Cient\'ificas y T\'ecnicas (CONICET), Godoy Cruz 2290, 1425,  Buenos Aires, Argentina}
\affiliation{Departamento de F\'isica, Facultad de Ciencias Exactas,  Universidad Nacional de La Plata,  1900, La Plata, Buenos Aires,  Argentina}

\affiliation{Instituto de F\'isica de L\'iquidos y Sistemas Biol\'ogicos (IFLySiB-CONICET) Universidad Nacional de La Plata, 1900, La Plata, Buenos Aires, Argentina}

\affiliation{Istituto dei Sistemi Complessi, Consiglio Nazionale delle Ricerche, via dei Taurini 19, 00185 Rome, Italy}

\author{Sergio A. Cannas}
 \affiliation{Consejo Nacional de Investigaciones Cient\'ificas y T\'ecnicas (CONICET), Godoy Cruz 2290, 1425,  Buenos Aires, Argentina}
\affiliation{Instituto de F\'isica Enrique Gaviola (IFEG-CONICET), Facultad de Matem\'atica Astronom\'ia F\'isica y Computaci\'on, Universidad Nacional de C\'ordoba, 5000, C\'ordoba, Argentina.}

\author{Wolfgang Losert}
\affiliation{College of Computer, Mathematical, and  Natural  Sciences, University of  Maryland, College Park, MD, USA}

\author{Dante R. Chialvo}
\affiliation{Instituto de Ciencias F\'isicas (ICIFI-CONICET), Center for Complex Systems and Brain Sciences (CEMSC3), Escuela de Ciencia y Tecnología, Universidad Nacional de Gral. San Martín, Campus Miguelete, 25 de Mayo y Francia,  1650, San Martín, Buenos Aires, Argentina}

\affiliation{Consejo Nacional de Investigaciones Cient\'ificas y T\'ecnicas (CONICET), Godoy Cruz 2290, 1425,  Buenos Aires, Argentina}

\begin{abstract}

In this article,  a correlation metric $\kappa_c$ is proposed for the inference of the dynamical state of neuronal networks. $\kappa_C$  is computed from  the scaling of the  correlation length  with the size of the observation region, which shows qualitatively different behavior near and away from the critical point of a continuous phase transition.  The implementation is first studied on a neuronal network model, where the results of this new metric coincide with those obtained from neuronal avalanche analysis,   thus characterizing well the critical state of the network.
  The approach is further tested with brain optogenetic recordings in behaving mice from a publicly available database.
Potential applications and limitations for its use with currently available optical imaging  techniques are  discussed.
\end{abstract}


\maketitle

The study of critical phenomena in the brain \cite{bak, chialvo2004critical,chialvo2010emergent} benefited from different experimental approaches. The most common by far is the statistical characterization of  the so-called  neuronal avalanches, consisting of sudden increases in the activity which exhibits power-law distribution of sizes and durations \cite{BeggsYPlenz}. This analysis has been reproduced over different setups (i.e., tissues and experimental conditions, see e.g. \cite{Mono,Rat}), and in a diversity of numerical simulations. The resulting statistics represent a long-term average estimation over thousands of avalanches, spanning very long periods of time, making the approach unsuitable for tracking fast dynamical changes. Several caveats, such as subsampling \cite{Tiagotesis}, thresholding \cite{AvaThreshold}, or the artifacts introduced by the coexistence of overlapping avalanches \cite{Korchinski}, as well as  alternative interpretations of the results  \cite{Destexhe} prompted the exploration of complementary approaches.

One of them, which is very often documented on continuous phase transitions, is the behavior of the correlation length $\xi$, which diverges with the size of the system {at the critical point} (see e.g. \cite{Cardy}), a fact that was shown to be exhibited by the large scale brain dynamics \cite{FraimanChialvo2012,Haimovici2013}. More recently  the same divergence of $\xi$ was demonstrated in the behaving mice brain \cite{tiago2020, Camargo}. These measures were facilitated by the use of novel optogenetic techniques \cite{Emiliani2015}, which allows for the recording of the individual activity of a relatively large number of neurons. In that work, a proxy of the standard finite size analysis, named box-scaling  was used, \cite{BoxScaling} in which the observation window, instead of the system size, is varied. An estimate of the correlation length $\xi$ was found to grow linearly or logarithmically with window size depending if the system is near or far from the critical state, respectively.
\begin{figure}[ht!]
\centering
\includegraphics [width = 0.99\linewidth] {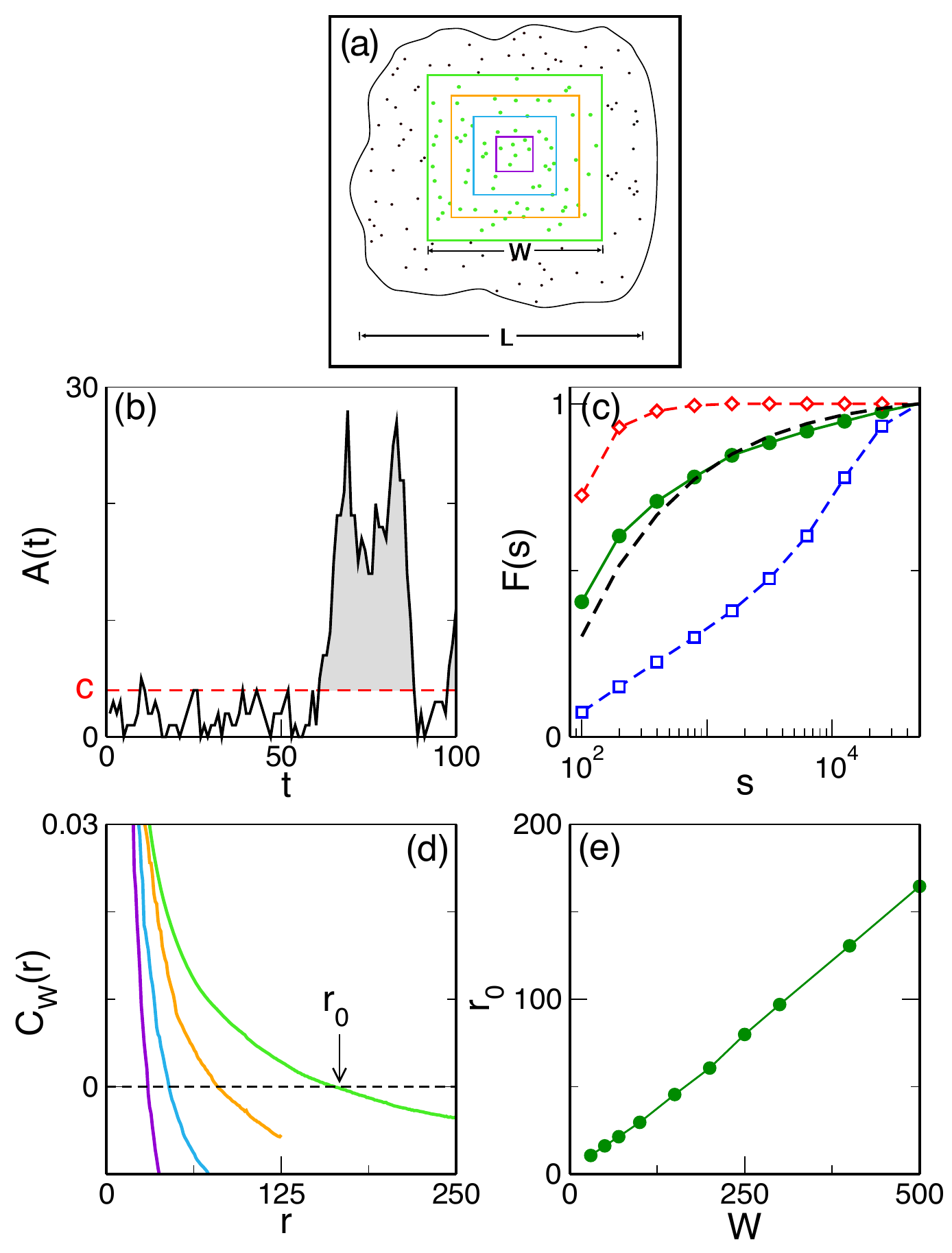}
\caption{System scheme. (a) A system of characteristic size $L$ is studied through boxes of side $W$. Only  neurons inside the box are recorded. (b) Example of the time series of $A$, the total number of active neurons inside a window, as a function of time. An avalanche, (filled with gray),  is defined as the total activity above a threshold $c$, computed from the time at which $A$ becomes greater than $c$ to the next time that it is becomes lower than $c$. Panel (c) shows the cumulative avalanche size distribution function $F(s)$ as a function of avalanche size $s$, for three different situations: subcritical ($T=0.33$, open blue squares), supercritical ($T=0.31$, open red diamonds) and close to criticality ($T=0.318$ green filled circles).  The dashed line represents the theoretical expectation for the avalanche size distribution expected at criticality{, $F^{NA}(s)$. Curves where computed for $m=10$ values of $s$}.  (d) The connected correlation function of a window of size $W$, $C_W(r)$, for several values of $W$,  computed  at criticality. From left to right, $W=50$ (violet line), $W=150$ (cyan),  $W=250$ (orange), $W=500$ (light green). The characteristic length $r_0$ for $W=500$ is marked with an arrow, as an example. (e)   Characteristic length $r_0$ as a function of window size $W$ at the critical state. Results computed on a system of size $L=1000$, $k=24$, $\pi=0.01$ and $T=0.318$. In panels (b) and (c), window size $W=500$ was used. }
\label{Fig1}
\end{figure}
Based on these previous results, the purpose of this letter is to introduce a simple metric,  describing the typical finite-size behavior of the  correlation length near criticality to distinguish critical from non-critical dynamics. To this end,  we study a simple model of neuronal dynamics that can be tuned towards and away from the critical point of a second-order phase transition dynamics, as the control parameter is varied. We contrast the new metric with the most common analysis, the avalanche size distribution  statistics.

The paper is organized as follows: Next we describe the model and define the observables, first for the standard metric of avalanches analysis and then for the finite-size correlation based metric. After that, the main results are described by contrasting the metrics in both numerical and experimental data. The paper closes with a short discussion of the limitations and potential applications.

\emph{Model and observables}- The model, previously described \cite{Haimovici2013,Zarepour,BoxScaling},  is a cellular automata  based on  the Greenberg and Hastings model \cite{Greenberg},  running on a two dimensional lattice of $L\times L$ neurons under periodic boundary conditions.  Each neuron $j$   has $k=24$ output connections chosen as follows: the closest $k$ neurons are initially  connected, and then, to mimic a small world topology, each of these connections is rewired with probability $\pi=0.01$ to another, randomly chosen, postsynaptic neuron within the whole system.  The resulting $k$ nonzero connection weights are taken randomly from an exponential distribution $p(W_{ij}=w)\propto exp(-w\lambda)$ with $\lambda=12.5$. (as in \cite{Zarepour}). The connection matrix is fixed and does not need to be symmetric. Time is considered discrete and each neuron $i$ may be in any of the following three states: quiescent ($S_i(t)=0$), active ($S_i(t)=1$) or refractory ($S_i(t)=2$). At time $t+1$  a quiescent neuron can become  active due to an external input with a small probability $r_1$ (we have used  $r_1=10^{-5}$), or if the  contribution of all active connections  at time $t$ is larger than a threshold $T$ ($\sum_j W_{ij} \delta_{S_j(t),1}>T$); an active neuron will became refractory always, and a refractory neuron will become quiescent with probability $r_2$ (following \cite{Zarepour}, we have used $r_2=0.3$ throughout the text). {The computer codes for numerical simulations and data analyses can be found in \cite{Codes}.} An important remark is that the results rely on universal behavior of the correlation function in critical phenomena, thus they are  model independent.

 We run simulations for several values of the control parameter $T$ which previous results \cite{BoxScaling} indicate produces subcritical (for very high values of $T$), supercritical (for very low values of $T$) or critical dynamics. To accumulate enough statistics, we run 20  numerical simulations (lasting $10^5$ time steps, discarding the initial 5000 time steps). For each simulation we constructed a different network with the same parameters $k$ and $\pi$ (i.e., the networks are stochastic realizations {each with different randomly chosen rewired connections and connection weights}). To mimic experimentally relevant situations, we record the dynamics of the neurons within a square window size of $W \times W$ neurons (with $W\leq L$), see Fig. \ref{Fig1}a. 

\emph{Metric based on the avalanche's size distribution}- The standard procedure for avalanche analysis \cite{BeggsYPlenz} focuses on  the estimation of the distribution of avalanche size and duration. For that, the total activity of the neurons inside a given  (spatial) window is  computed as a function of time, $A(t)=\sum_{i \in W\times W} \delta_{S_i(t),1}$  ($S_i(t)=1$ if neuron $i$ is spiking at time $t$). Notice that in the standard procedure it is usual to group the activity on time bins approximately equal to the average of  all
inter-spike-intervals. The coarse grain scale of the model considered here (i.e., only three discrete states) determines that we must compute $A(t)$ for each time unit, as mentioned above.  Also, since for the conditions in our case $A(t)$ very rarely  becomes zero, following \cite{AvaThreshold}, we need to define a non-zero avalanche threshold $c$. Avalanche size {$s$} is defined then as the total activity above $c$ between two consecutive zeros of $A(t)-c$ {(i.e., $s= \sum_{t}[A(t)-c]$, where the sum is performed over the avalanche duration)}, see Fig. \ref{Fig1}b. At criticality, avalanche size distribution, $P(s)$, is expected to have a power law distribution, $P(s) \propto s^{-\tau}$, where, in the mean field directed percolation universality class, $\tau=3/2$ \cite{BeggsYPlenz,Zapperi}.
The value of $c$ is chosen to maximize the number of avalanches for each value of $T$ and $W$.

The goodness of fit of the neuronal avalanches size distribution to a power law has been considered as suggestive for critical dynamics, which taken in isolation may call for caveats, precautions, and criticisms \cite{PowerLaw}. Nonetheless, when used in conjunction with other measures it can overcome some of its limitations \cite{Destexhe}. In that regard, Shew \emph{et al.} \cite{KappaShew} defined, from the {observed} cumulative  avalanche size distribution, $F(s)$, a metric $\kappa_S$, which is \cite{KappaShew}: 
\begin{equation}
 \kappa_{S}= 1+{1\over  m} \sum_{k=1}^{m} F^{NA}(\beta_k)-F(\beta_k)\label{EqBeta}
\end{equation}

Where $F^{NA}(\beta)={ 1-(s_{min}/\beta)^{\tau-1} \over  1-(s_{min}/s_{max})^{\tau-1}}$, is the theoretical distribution for the critical case, and $\beta_k$ are $m$  logarithmically spaced values ranging from $s_{min}=50$ to $s_{max}=50 000$. We have used $m=10$ as in \cite{KappaShew}. {An example of $F(s)$ and $F^{NA}$ is shown in Fig. \ref{Fig1}c.}
  For {power law} avalanche size distributions with exponent $\tau=3/2$, { the cumulative avalanche size distribution $F(s)$ will be equal to $F^{NA}(s)$, then}  a value of $\kappa_{S}=1$ is expected, while $\kappa_{S}\gtrless 1$ for super/subcritical conditions. 

\begin{figure}[ht]
\centering
 \includegraphics [width = 0.98\columnwidth] {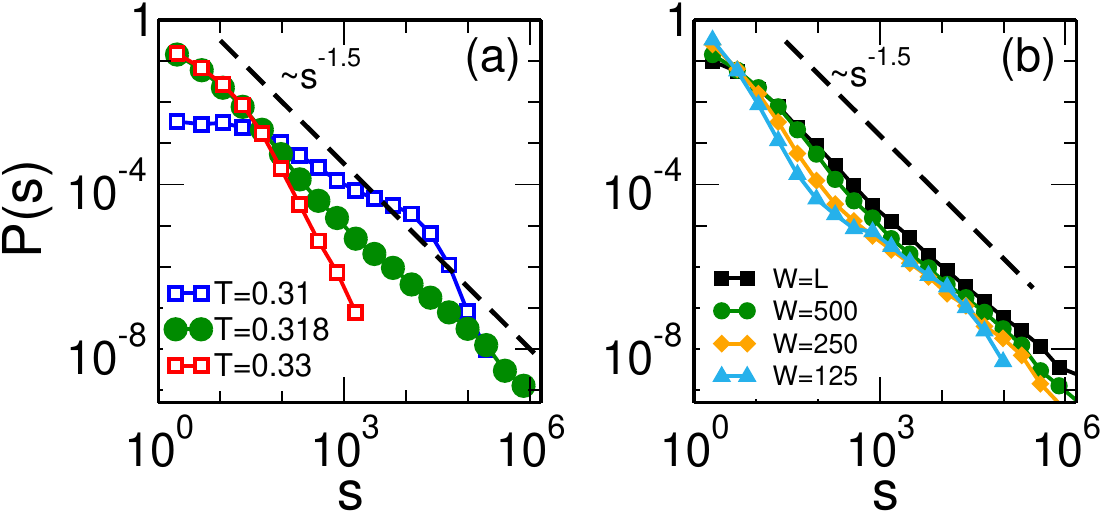}
 \caption{Avalanche size distribution computed on a window of size $W=500$, for different values of $T$ in (a) and for $T=0.3180\simeq T_C$ and several values of $W$ in (b). The dashed lines, in both panels, show a power law with exponent $-3/2$ as a guide to the eye. All parameters  as in Fig. \ref{Fig1}.}
\label{Fig2}
\end{figure}

\emph{Metric based on finite-size scaling of correlations}- Following previous work \cite{BoxScaling}, we computed the connected correlation function on a window of size $W$, as the correlation of the fluctuations of the neuronal activity,  with respect to it's \emph{instantaneous spatial average} \cite{cavagna2010,FraimanChialvo2012,tang,tang2,cavagna2014,Haimovici2013,flocks,tiago2020,BoxScaling,grigera,MarianiCorrLength}:
\begin{equation}
C_W(r)= {1 \over c_0} { \sum_{i,j}  \delta v_i \delta v_j \delta(r-r_{ij})  \over \sum_{i,j}  \delta(r-r_{ij})  } 
\label{eq_box}
\end{equation}
where $\delta(r-r_{ij})$  is a smoothed Dirac $\delta$ function selecting pairs of  neuron states at a distance $r$ (in practice, we have computed $C_W(r)$ for integer values of $r$, averaging all points at distances $(r-0.5,r+0.5]$); $r_{ij}$ is the Euclidean distance from the site $i$ to site $j$;  $\delta v_i$ is the value of the signal $v_i$ of site $i$ at time $t$, after subtracting  the \emph{instantaneous spatial average} of signals $V(t)= (1/N)\sum_i^N v_i(t)$, i.e., $\delta v_i(t)=v_i(t)-V(t)$; and ${1 \over c_0}$ is a normalization factor to ensure that $C_W(r=0)=1$.  We consider that $v_i=1$ if neuron $i$ is in the active ($ {S}_i=1$) or refractory ($ {S}_i=2$) state and $v_i=0$ otherwise.  {Although $C_W(r)$ can be computed on a single snapshot (in contrast with $F(s)$), to improve statistics, we average the result over several time steps.} We compute Eq. \ref{eq_box} once every 20 time steps  (i.e., we take information for 
4750 \cite{Comment4750}
time steps for each network), and then average the result over different time steps and different networks. An estimate of the  correlation length can be calculated from Eq. \ref{eq_box} as $r_0$, the first zero crossing of the function (i.e., $C_W(r_0)=0$). {An example of $C_W(r)$, for different values of $W$, is shown in Fig. \ref{Fig1}d, while $r_0$ as a function of $W$, is shown in Fig. \ref{Fig1}e.} We remark that the implementation of $r_0$  estimates correlations  computed  inside  a window, after subtracting the \emph{instantaneous} spatial average. This differs from the frequently considered connected correlation function, computed from the fluctuations of each variable with respect to their time average (although, for systems in equilibrium thermodynamics, they are equivalent \cite{flocks}). This characteristic makes $C_W(r)$ in Eq. \ref{eq_box}, immune to global trends and hidden confounders as discussed elsewhere \cite{BoxScaling,grigera}.

We measure $C_W(r)$  for several values of $W$, ranging from $W_{min}$ to $W_{max}$. For equilibrium thermodynamic systems, the behavior of $r_0$ as a function of $W$, for fixed $L$,  is known in the limiting cases: $r_0\propto W$ for $W\ll L\ll \xi$ at criticality, while  $r_0\propto \xi \log(W/\xi)$ for $\xi\ll W_{min}$, where $\xi$ is the standard correlation length, see \cite{flocks,BoxScaling}.

To estimate the distance to criticality, for each explored window size $W_i$, we propose a linear relation among $r_0(W_i)$ and $W_i$, $r_0(W_i)=a_i \times W_i$, and extract the value of the slope $a_i$ from the data \cite{CommentExp}. Also, we propose a 
logarithmic growth $r_0(W_i)=r_0(W_{min})+b_i \log(W/W_{min})$.
Similar to Eq. \ref{EqBeta}, we define
\begin{equation}
 \kappa_{C}= {CV_s^2 \over CV_c^2+CV_s^2},\label{EqD}
\end{equation}
where $CV_s$ is the coefficient of variation of $\{b_i\}$, and $CV_c$ is the coefficient of variation of  $\{a_i\}$
( see \cite{Codes}). Notice that $0\leq \kappa_{C} \leq 1$, where  $ \kappa_{C}=0$ is for a perfect logarithmic growth and $ \kappa_{C}=1$ is for perfect linear growth. While more sophisticated measures can be proposed,  the definition of (\ref{EqD}) is simple and insensitive to changes in the spatial scale ($r\to \lambda r$).

\begin{figure}[ht]
\centering
 \includegraphics [width = 0.98 \linewidth] {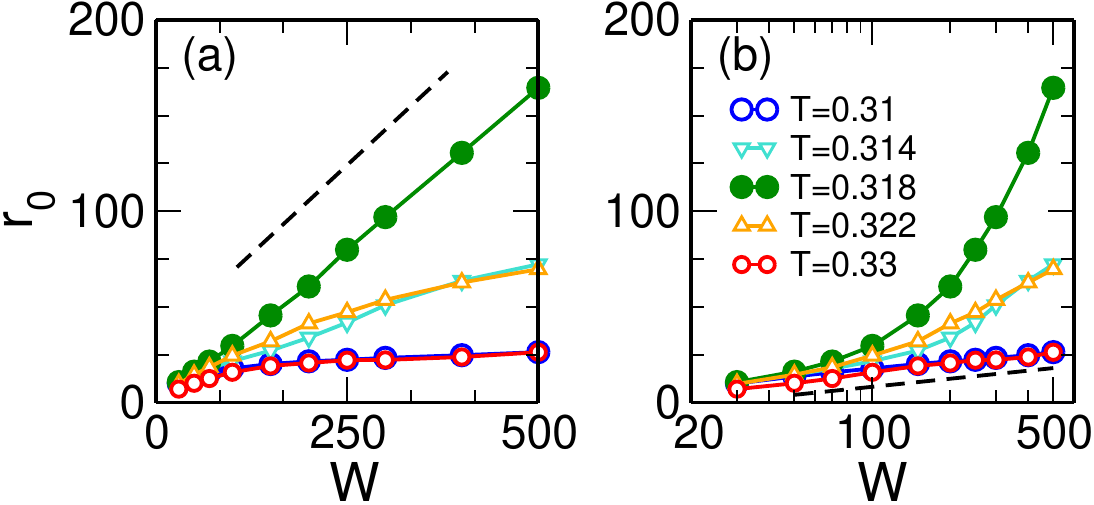}
 \caption{Characteristic correlation length as a function of window size $W$ obtained at various  control parameter values $T$ (indicated in the legend). The same results are plotted in linear scale in panel (a) and in linear-logarithmic  scale in panel (b).  All other parameters  as in Fig. \ref{Fig1}.}
\label{Fig3}
\end{figure}
%

\emph{Results}- As a reference, we first characterize the behavior of the avalanche  size distribution, computed inside of a window of size $W=500$, for different values of $T$. The results are  shown in Fig. \ref{Fig2}a. In the subcritical state ($T=0.33$), activity is low, and there are no large avalanches, for any value of $c$. In the supercritical case ($T=0.31$), activity is very high, being always larger than zero.  The avalanche size distribution has a hump for $s\sim 10^5$. Hump position depends on $c$, showing system-wide avalanches (commonly dubbed ``dragon kings'') for low values of $c$. In the critical case  ($T\simeq 0.318$), avalanche size distribution follows closely a power law with exponent $\tau=3/2$. Different values of $\tau$, in the range [1.3-1.7], can be estimated for different values of $c$.  For the critical data in the figure (line with  circles in Fig. \ref{Fig2}a), it can be seen that for small values of $s$ (i.e., $s<100$), there is an excess of  avalanches, compared to the expected. This excess is a consequence of subsampling, and is not present for $W=L$, while it is even larger for small values of $W$, such as $W=125$, see Fig. \ref{Fig2}b. This difference  may be due to the contributions of avalanches that enter or leave the window from the rest of the system, as already discussed in the context of avalanches in the qKPZ model, see Ref. \cite{ChenSethna}. 

Next, we turn to describe the correlation behavior on the same data used to study avalanches. The characteristic correlation length $r_0$ as a function of window size $W$, for $W_{min}=30$, $W_{max}=500$, is shown in Fig. \ref{Fig3}. For the critical value of the threshold ($T=0.318$), there is a linear relation among $r_0$ and $W$, while for sub and supercritical regimes, $r_0$ is smaller, and the growth of $r_0$ with $W$ is logarithmic. Slightly subcritical and supercritical cases, (plotted  with triangles), show intermediate results.  Similar results can be found when $C_W(r)$, Eq. \ref{eq_box}, is computed for the whole system ($W=L$), varying system size, as shown in \cite{BoxScaling} for the Ising paramagnetic-ferromagnetic model and for a different neuronal model \cite{Tiagotesis}.

\begin{figure}[ht!]
\centering
 \includegraphics [width = 0.95 \linewidth] {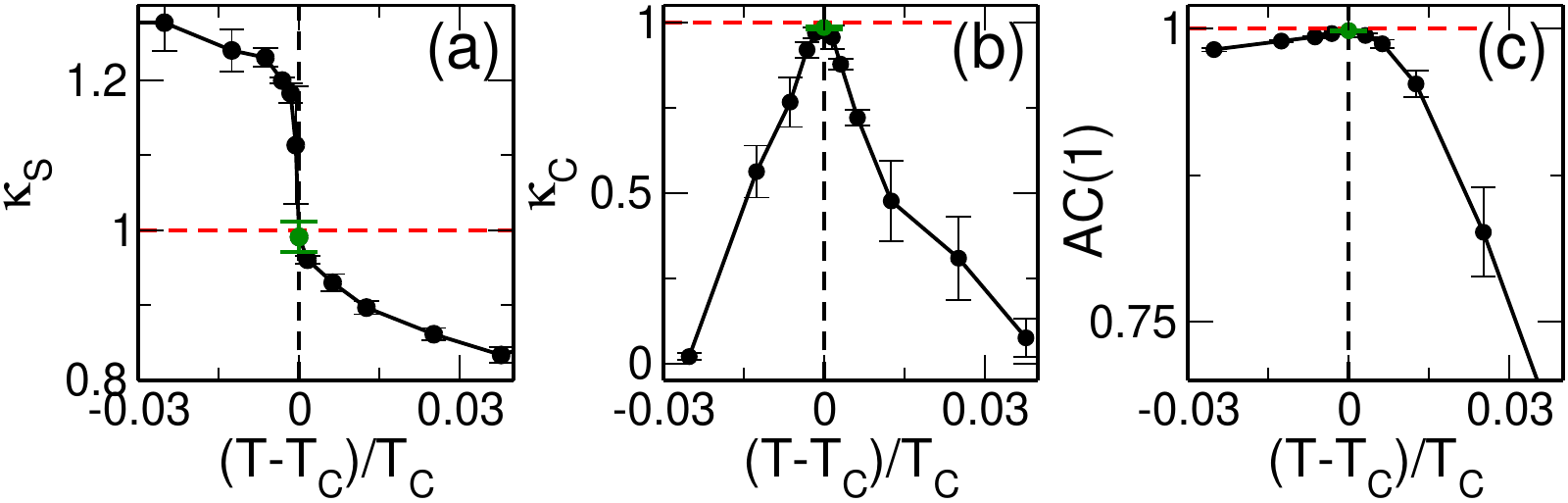}
 \caption{Behavior of the  different metrics (mean $\pm SD$) as a function of $T$ near the critical point of the neural model: $\kappa_{S}$ in panel (a) and $\kappa_C$ in panel (b) and  AC(1) in panel (c).  All other parameters as in Fig. \ref{Fig2} and \ref{Fig3}.}
\label{Fig4}
\end{figure}

The values of $\kappa_{S}$ and $\kappa_C$, extracted from avalanche size distribution and correlation length scaling, are shown in Fig. \ref{Fig4}. Avalanche analysis ($\kappa_S$), assuming $\tau=3/2$,  yields expected results: $\kappa_S > 1$ ($<1$) for supercritical (subcritical) regime, while $\kappa_S$ is closest to $1$ for critical regime, $T=0.318$ (marked with a green dot). For very subcritical values (high $T$), $\kappa_S$ does not keep on decreasing, probably due to having a short range of $s$ values captured by $P(s)$. The analysis of characteristic length collapse,  $\kappa_C$, shows compatible results, see Fig. \ref{Fig4}b. The linear fit is better that the logarithmic fit (i.e., $\kappa_C>0.5$) only for $0.314<T<0.322$, having its peak at $T=0.318$, i.e, the same value as in $\kappa_S$.  

For completeness, in Fig. \ref{Fig4}c we also show the first autocorrelation coefficient of the activity, $AC(1)$ which by definition is always smaller than 1, and reaches a maximum at criticality \cite{Control}. $AC(\Delta t)$ is computed from the activity $A(t)$ on the largest window ($W=500$) as $$ AC(\Delta t)=\langle A(t+\Delta t)- \langle A\rangle \rangle \times \langle A(t)- \langle A\rangle \rangle/[\langle A(t)^2\rangle - \langle A(t)\rangle^2] $$, where $\langle ...\rangle$ stands for temporal average.  The critical value of $T$ derived from {$AC(1) = AC(\Delta t=1)$}  also coincides with results from $\kappa_C$ and $\kappa_S$.

\begin{figure}[ht!]
\centering
 \includegraphics [width = 0.99 \linewidth] {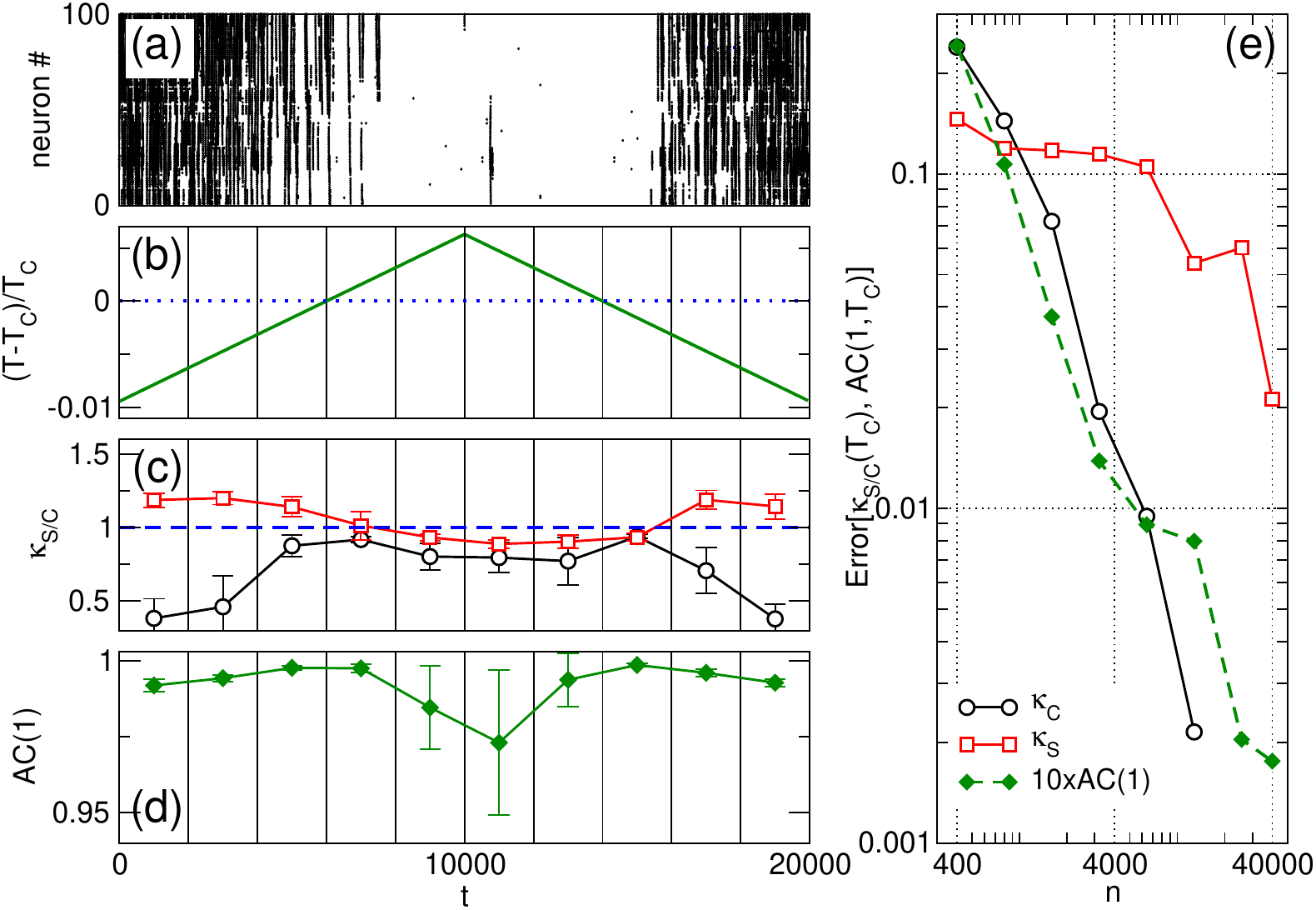}
 \caption{Numerical simulations demonstrating the behavior of the metrics in response to a slow change in the network excitability, here simulated by ramping up and down the model parameter $T$. Panel (a) shows the raster plot of a subset of 100 neurons as $T$ is varied. Panel (b) shows the evolution of $T$ as a function of time $t$.  Panel (c) shows the estimated mean (+- SD) $\kappa_S$ and $\kappa_C$, computed over time segments of $n=2000$ steps. For $\kappa_C$, we used spatial windows $W\leq 300$. For $\kappa_S$ an average of  $\sim$ 95 avalanches (range 14-280) were detected in each run and each temporal window. Panel (d) shows the first autocorrelation coefficient $AC(1)$ of the population rate fluctuations within the same windows.
Panel (e) shows the errors of the estimators, computed as the average distance between the measured and the expected value, $\kappa_{S/C}^*$ and $AC(1)^*$, as a function of the number of steps $n$, at $T=0.3180\simeq T_C$. 
Results in each panel are from four independent realizations of the numerical simulations. For avalanche analysis, since $n$ is variable, we considered $s_{min}$ as 10 times the smaller avalanche size observed, and $s_{max}$ as 0.1 of the largest observed avalanche size. }
\label{Fig5}
\end{figure}

To compare the performance of $\kappa_S$, $\kappa_C$ and $AC(1)$, in Fig. \ref{Fig5}, we show the metric's behavior as a function of slow variations of the control parameter $T$. 
{In Fig \ref{Fig5}b, we show how the control parameter $T$ is varied as a function of time, generating a non-stationary activity time-series (see  the raster plot for a few neurons in panel a)}. 
The values of $\kappa_S$, $\kappa_C$ and $AC(1)$, computed on time segments of $n=2000$ steps, are shown in panels c and d.  It can be seen that close to criticality (i.e., $T=T_C$), the variability in $\kappa_C$ is lower than the variability in $\kappa_S$. We also show the first autocorrelation coefficient of the activity, $AC(1)$  (see Fig. \ref{Fig5}d), which shows a  low variability  in the  critical (and supercritical) regime.

To study this observation in depth, we run four independent  simulations on the same network  (with different annealed noise), at fixed $T=T_C$, for 40 000 steps each. Using all this data (i.e., all the time frames from all the runs), we compute the expected values $\kappa_{S/C}^*$ and $AC(1)^*$. Next,  we compute    $\kappa_{S/C}$ and $AC(1)$ using several short segments of the time series,  of length $n$ (from $n=400$ to $n=40 000$). We define the \emph{Error} as the average distance (computed as the absolute difference) of these values to the expected values  $\kappa_{S/C}^*$ and $AC(1)^*$. For all  observables, the \emph{Error} decays with  the number of samples used $n$ (see Fig. \ref{Fig5}e). For samples with $n>1000$, we find that the  \emph{Error} in $\kappa_C$ (and $AC(1)$) is lower than the  \emph{Error} in $\kappa_S$. More important, the error of $\kappa_C$ and $AC(1)$   decay as $\sim 1/n$, faster than for  $\kappa_S$.

\begin{figure}[hb!]
\centering
\includegraphics [width = 0.9\linewidth]{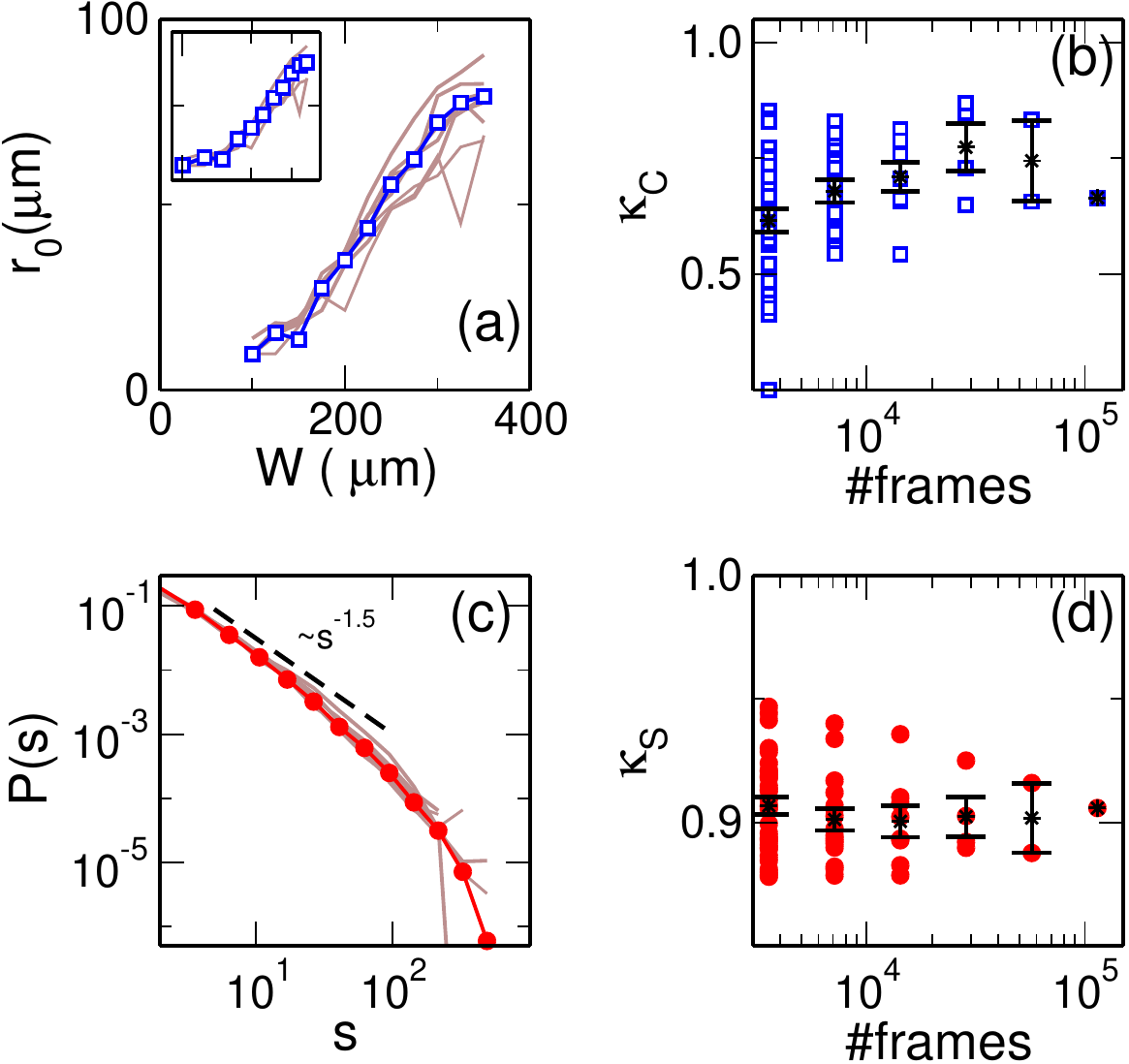}
\caption{Experimental recordings. Analysis of neuronal spike data inferred from two-photon imaging  from the Allen Institute database \cite{Allen}. Panel (a) shows box-scaling results for the entire data with empty symbols and  with lines for the segments of 1/8 of the time series. The inset shows the same data in log-linear axis. In panel (b) the symbols show $\kappa_C$ for different time windows, as a function of the number of frames used, while the black lines show the mean $\pm$ standard error. 
 Panel (c) shows the avalanche size distribution of  the same data, for all time frames with filled symbols  (17155 avalanches), and for segments of 1/8 of the points (as in Panel (a)) with lines.  
 Panel (d) shows the $\kappa_S$  in the same format used in Panel (b). For each time segment, $\kappa_C$ was computed from windows of size 100 $\mu m$ or larger, while $\kappa_S$ was computed using avalanche sizes ranging from twice the minimum observed avalanche size, to  half of the largest observed avalanche size. }
\label{Fig6}
\end{figure}

\begin{figure}[ht!]
\centering
\includegraphics [width = 0.99\linewidth]{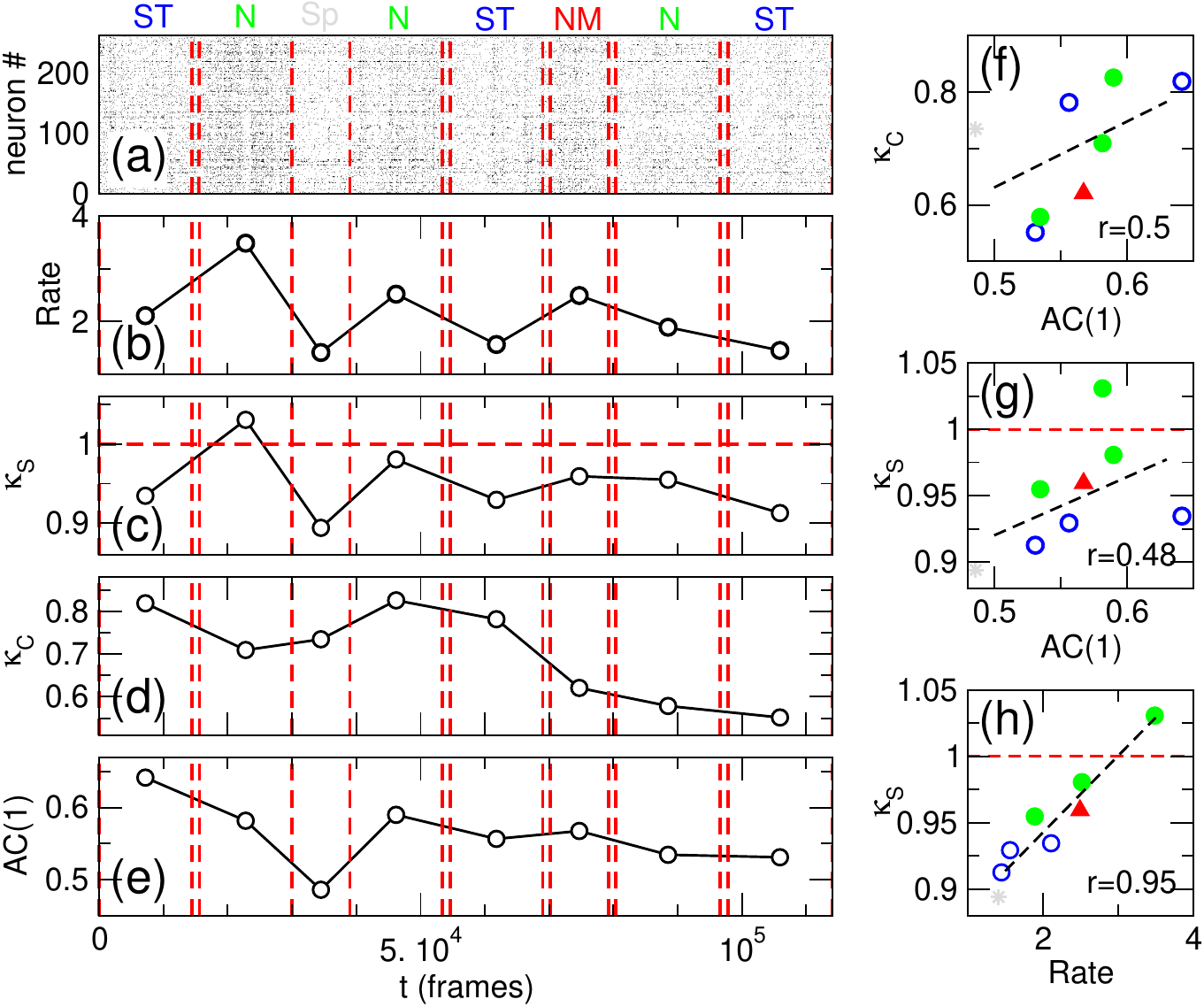}
 \caption{Dynamical changes in response to visual stimuli exhibited by the experimental recordings according with the different metrics (same data as in Fig. \ref{Fig6}). Panel (a) shows the raster plot, and panel (b) the average rate (number of spikes per frame) computed over time blocks related to different visual stimuli. Panels (c) through (e) show  $\kappa_S$, $\kappa_C$, and  $AC(1)$ respectively, for the same time blocks. These points are replotted in the right panels where Panel (f) shows $\kappa_C$ as a function of $AC(1)$ and
 panels (g)-(h) show $\kappa_S$ as a function of $AC(1)$  and as a function of the Rate, respectively ($r$ values correspond to linear regression coefficients). 
The  visual stimuli, labeled in panel (a), and denoted by the vertical dashed lines, consisted of a sequence of 8 min of static gratings (ST)  followed by inter-stimulation period of gray screen, 8 min of natural images (N), 5 min of spontaneous activity (Sp), 8 min. of natural images, inter-stim gray screen, 8 min. of static gratings, inter-stim gray screen, 5 min of natural movie (NM), 9 min. of natural images, and 9 min of static gratings. The  symbols in panels (f)-(h) correspond to the different stimuli kinds  (colored as the labels on top of panel (a)). Results computed from \cite{Allen}.}
\label{Fig7}
\end{figure}

Novel optogenetic imaging techniques allow for the simultaneous recordings of the activity of hundreds of neurons \cite{Emiliani2015}, an optimal setting to compare the statistical measures.
Figs \ref{Fig6} and \ref{Fig7} show the behavior of the proposed metrics to characterize the dynamics of a selected dataset from the Allen Institute's Brain Observatory \cite{Allen}, recorded (at 30Hz for 114099 time frames) from a conscious mouse. The data  corresponds to the inferred spike probabilities of 295 neurons inside a field of view of  400 $\times$ 400$\mu m$ in the VISp area. This data set was selected because its experimental design includes the presentation of different visual stimuli. We expected that the stimuli shall induce variations on the neuronal network state large enough to be reflected consistently on the metrics described here. {Previous analyses on rat visual cortex \cite{RatStim}, subject to monocular deprivation, and turtles subject to visual stimulation  \cite{TurtleStim} showed that the stimulation produced changes on the dynamical state, that were measurable using computations related to avalanche size distribution and other proposed observables. 
} 

First we  explored the behavior of the metrics as a function of the number of samples (i.e., frames). Fig. \ref{Fig6}a shows the box-scaling results, calculated from the spike time series extracted from \cite{Allen}. A linear relation between $W$ and $r_0$ is observed for all $W > \sim 100 \mu$m, while this relation breaks at shorter distances.  
From this observation, we estimate a characteristic interaction length of the order of $100\mu m$, which is  slightly shorter than (but comparable to) the experimental neuronal connection lengths \cite{Distance} 
The same data is plotted in log-linear axis in the inset of that figure, to emphasize its non-logarithmic scaling (compare with results in Fig. \ref{Fig3}). Fig. \ref{Fig6}c shows the avalanche size distribution, computed from the same spike time series. The results approximate the expected power law distribution for about two decades.
The values of $\kappa_C$ and $\kappa_S$, for different sampling length, are shown in Fig. \ref{Fig6} (b) and \ref{Fig6} (d). Note that, as expected, the range of $\kappa_C$ and $\kappa_S$ observed values broaden for shorter time series.

Next, we explored up to which degree  the fluctuations, spontaneous or introduced by the visual stimuli, may be reflected on the  proposed metrics. 
Fig. \ref{Fig7} shows the results of  analyzing the temporal fluctuations of the metrics computed in eight non-overlapping temporal segments, each one corresponding to different visual stimuli. According to the analysis,  throughout the segments the dynamics remain slightly subcritical, with variations depending on the type of stimulus. In consequence, the relative fluctuations of each metric are directly proportional to each other, as shown in  Fig. \ref{Fig7} (f)-(h).  Note that the population rate (i.e. panel (b)), in this context, shall be considered as a pseudo-order parameter \cite{Control}.

While inferring the dynamical state of the network is relevant on its own, another important question, in the context of brain dynamics, is how the dynamical state may affect the system's response. To address this question,
    we study how the neurons' response depends on the network state. We define the response to a given stimulus as the firing rate change when the stimulus is turned on, compared to the rate immediately before, divided by the summed rate: $Response={R_s-R_b \over R_s+R_b}$, where $R_s$ is the rate when the stimulus is present,  averaged over all considered neurons and stimulus presentations, and $R_b$ is computed over the same neurons, for time windows of the same duration, immediately before the stimulus onset. For static gratings, we say that a neuron responds to a given angle if the response is larger for  that orientation than for gratings in any other direction. 
Similarly, we say that a neuron responds to a given natural image if the rate increase is larger for that image than for any other natural image.

\begin{figure}[ht!]
\centering
\includegraphics [width = 0.99\linewidth]{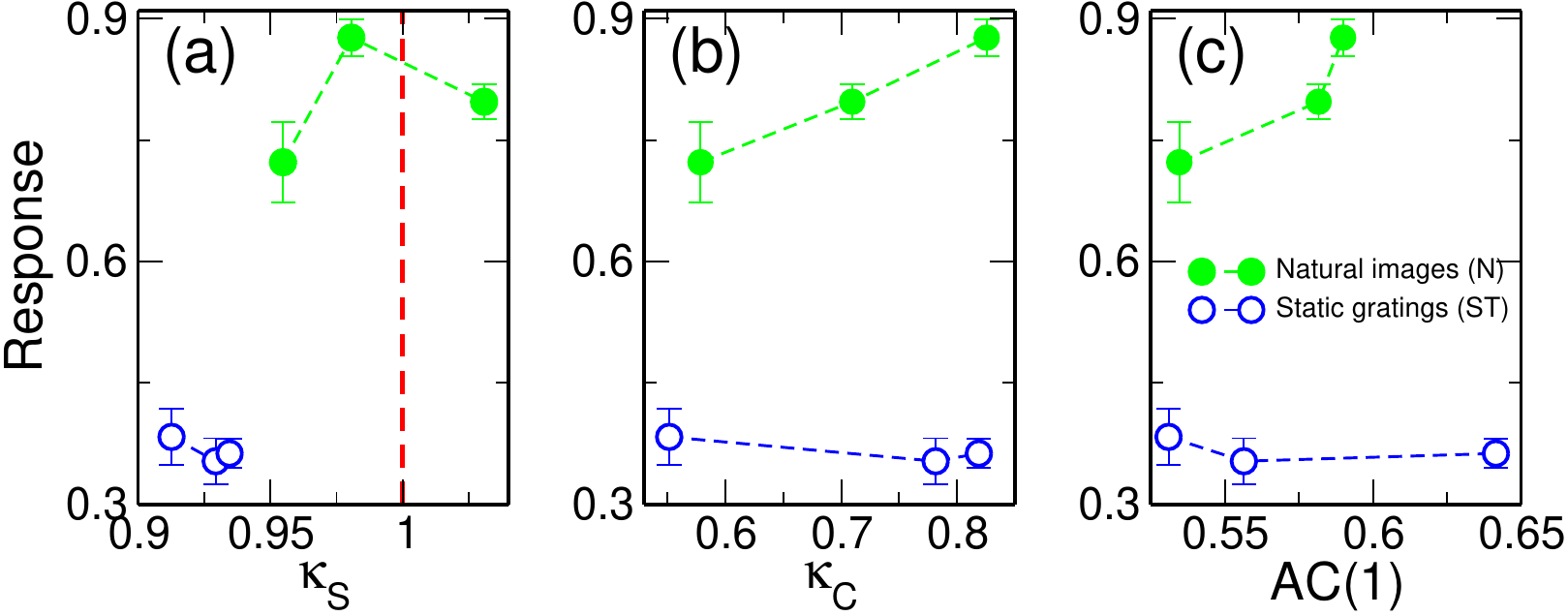}
 \caption{Changes in the level of response to natural images (N) and static gratings (ST) stimuli as a function of the network state, estimated by the three metrics. Panel (a) corresponds to $\kappa_S$, panel (b) to $\kappa_C$  and  (c) to $AC(1) $. For static gratings, the response was computed on the 5 most responsive neurons on each orientation (total: 30 neurons). For natural images, the neurons for the 8 natural images with most responsive neurons were considered (natural images 45 with 20 responsive neurons; 85, with 14 neurons; 41, 115, and 108, with 10 neurons; 69 and 36, with 8 neurons; 86 with 7 neurons, total: 87 neurons). All other parameters are as in Fig. \ref{Fig7}.}
\label{Fig8}
\end{figure}

Fig. \ref{Fig8} shows the change in network responses for different network states, evaluated with different metrics. The results show that  the \emph{response} for static gratings  is mostly insensitive to  the changes in the dynamical state, while the response for natural images became larger when the state approaches criticality. We have limited the analysis to the 8 natural images that generate the largest responses. The analysis is a pilot demonstration of two aspects that deserve to be better explored: on one side it shows the well-known fact that the response of the visual cortex is stronger for natural images, and on the other side, that when the metric indicates that the network is closer to criticality it maximizes its responses \cite{KappaShew}.

 {\emph{Discussion-} 
It is known that the status of cortical networks changes following spontaneous fluctuations in excitability, arousal, sleep, vigilance or in response to sensory inputs or anesthetic agents.  A simple approach to track these changes is the computation of the pair-wise mutual correlations, which at the critical state exhibits scale-invariance.  A motivation for  the present work is to develop practical methods for tracking these changes in the global correlations of a network, under the assumption that such quantification may help to understand cortical responses under a variety of changing circumstances.    

New methods shall take advantage of the novel optogenetic techniques  which not only provide data from a very large number (hundreds to thousands) of  neurons  but also provide  spatial information. In contrast, the avalanche analysis only relies on counting the number of neurons firing at any given time, not profiting from the abundance of spatial information offered by optogenetic techniques.
The metric proposed here, based on the computation of the connected correlation length, is performed from instantaneous snapshots of the system. By construction, this feature gives the approach some important advantages, for instance more immunity to spurious collective effects (non-critical) from a trivial driving by a hidden variable. This cannot be singled out by standard avalanches analysis. In addition, box-scaling should not be affected by sub sampling artifacts \cite{Tiagotesis} or overlapping avalanches \cite{Korchinski}, since the value of $C_W(r)$ is computed from the activity of pairs of  \emph{observed} neurons at a distance $r$.  

Regarding the sensitivity of the different observables computed here, we should stress that the definition of $\kappa_S$, Eq. \ref{EqBeta}, computes the signed distance to the expected power law distribution (instead as, for example, the absolute distance), in such way that positive and negative deviations from the ideal cumulative distribution  (as seen  in the critical curve of Fig. \ref{Fig1}c, for low and high values of $s$) compensate. While this makes $\kappa_S$ robust in the absence of enough data, it also makes it less sensitive. Also, notice that the actual value of $\kappa_S$ depends on different parameters (such as bin length or threshold {$c$}).

AC(1) has a broad peak about the critical point,  which makes it an excellent observable for directing the system towards criticality, as discussed already in \cite{Control}. While this feature is shared with the $\kappa_C$  approach, the later requires much more information (i.e., to compute from all pairs). {Since both peak at the critical point, they cannot be used to distinguish subcritical from supercritical regimes, and some other observable, such as rate or $\kappa_S$ has to be used in conjunction to disambiguate. Nevertheless, it should be stressed that supercritical regimes are infrequent in neuronal data.}

    Notice that, similar to $\kappa_S$,  AC(1) is computed from the time series of the population activity, which means that it may be subject to external biases {and nonstationarities}, and they do not  profit from spatial information. On the other hand, $\kappa_C$ can be computed from single time frames, but it  cannot be calculated if the neurons' positions are unknown, or in systems where positions are ill-defined. 
    Also, as in the numerical results of \cite{tiago2020}, we have found in experimental data that the linear relation between $r_0$ and $W$ at criticality breaks down for very small windows, an observation that deserves further research efforts, and has to be taken into account if $\kappa_C$ is intended to be used on very small system sizes.

Overall, numerical simulation results show that the value of the control parameter $T_c$ (i.e., for critical behavior) inferred via avalanche-size distribution is very close to the value that maximizes the correlation length. Thus,  the long-term state of the system can be monitored from either method, although the computation of the correlation length  should be more sensitive to dynamic changes, and less dependent  on parameters. The analyzed experimental data support this picture. 

Many of the results on neuronal activity (including those studied here \cite{Allen}) on behaving animals are nowadays obtained from optogenetic recordings \cite{Emiliani2015}, in which the spike of a neuron (lasting about 1ms) generates an optical response, related to the displacement of calcium within the neuron, that decays on larger time scales (in the order of a few hundred of milliseconds).  Typically, neuronal spikes are inferred through the deconvolution of that signal. However, it has been recently proposed that some analyses, related to different kinds of  correlations among pairs of neurons,  may be performed without requiring a deconvolution  \cite{Behtash}. 
Although it is not the objective of the present work, the computation of $\kappa_C$  from  minimally pre-processed (i.e., normalized or  z-scored) calcium data yields results qualitatively similar to those presented above from the inferred spike data. This is a promising avenue for an approach that  does not depend on the intricacies of the deconvolution algorithms. The relation between $\kappa_C$ results obtained from raw calcium signals and from spike data  deserves further research, and would likely benefit from the analyses proposed in \cite{Behtash} (see also \cite{Behtash2}).

In summary, we have explored ways to  estimate changes in a network status and introduced a simple metric,  $\kappa_C$, describing the typical finite-size behavior of the (instantaneous) spatial correlations of neuronal activity. By  construction, $\kappa_C$ is able to distinguish critical from non-critical dynamics and  compares well  with avalanche analysis which estimates the distribution of the space-integrated activity. In a given experimental situation, the observation of large $\kappa_C$ values indicating long-range \emph{spatial} correlations is consistent with the simultaneous observation of large values for the \emph{temporal} correlations, as shown previously \cite{Control}. Results presented here suggest that the correlation length computations using box-scaling are well suited as a complement or a substitute of neuronal avalanche analysis as a useful tool for monitoring criticality  on diverse experimental conditions.

\emph{Acknowlegments:}    This work was partially supported by Grant No. 1U19NS107464-01 from NIH BRAIN Initiative (USA) and CONICET (Argentina).


\begin{thebibliography}{55}
 
 \bibitem {bak}    {P. Bak}, \emph {  How nature works: The science of self-organized criticality} (Springer Science, New York, 1996). 

 \bibitem {chialvo2004critical}   D. R. Chialvo, \href {https://doi.org/https://doi.org/10.1016/j.physa.2004.05.064}  {   Physica A \textbf  { 340}, {756} (2004)}.
  

 \bibitem {chialvo2010emergent}  D.~R. Chialvo, \href {https://doi.org/10.1038/nphys1803}  { Nat. Phys. \textbf  { 6}, {744} (2010)}. 
 
  \bibitem {BeggsYPlenz}    J.~M. Beggs and  D. Plenz, \href {https://doi.org/10.1523/JNEUROSCI.23-35-11167.2003}  { J. Neurosci.\textbf{ 23}, 11167 (2003)}. 

  \bibitem {Mono}  T. Petermann, T. C. Thiagarajan, M. A. Lebedev, M. A. L. Nicolelis, D. R. Chialvo and D. Plenz, \href {https://doi.org/10.1073/pnas.0904089106}{ Proc. Natl. Acad. Sci. USA \textbf{106}, 15921–15926.  (2009). }  
 
  \bibitem {Rat} N. Friedman, S. Ito, B. A. W. Brinkman, M. Shimono, R. E. Lee DeVille, K. A. Dahmen, J. M. Beggs and T. C. Butler,  \href {https://doi.org/10.1103/PhysRevLett.108.208102}{Phys. Rev. Lett. \textbf{108}, 208102 (2012).} 
 
 
 \bibitem {Tiagotesis}  { T.~L. Ribeiro,   S. Ribeiro,   H. Belchior,   F. Caixeta and   M. Copelli, \href {https://doi.org/10.1371/journal.pone.0094992}  {   PloS one \textbf  { 9}, {e94992} (2014)}}. 

 
 \bibitem{AvaThreshold} P. Villegas, S. di Santo, R. Burioni and M.A. Muñoz, \href {10.1103/PhysRevE.100.012133}{Phys. Rev. E \emph{100}, 012133 (2019).} 

 \bibitem{Korchinski} D. J. Korchinski, J. G. Orlandi, S.-W. Son and J. Davidsen, \href {10.1103/PhysRevX.11.021059}{
Phys. Rev. X 11, 021059 (2021)}  
 
 
 \bibitem{Destexhe} J. Touboul and A. Destexhe,  \href{10.1103/PhysRevE.95.012413  }{Phys. Rev. E \textbf{95}, 012413 (2017)}.  
 
 \bibitem{Cardy} J. L. Cardy,  (ed.) \emph{Finite-size Scaling} (North Holland, Amsterdam, 1988).
 
  \bibitem {Haimovici2013}    {A. Haimovici},  { E. Tagliazucchi},  { P. Balenzuela} and  { D.~R. Chialvo}, \href {https://doi.org/10.1103/PhysRevLett.110.178101}  {  { Phys. Rev. Lett.} \textbf  { 110}, {178101} (2013)}. 
  


  
   \bibitem {FraimanChialvo2012}  {  {D. Fraiman} and  { D. Chialvo}, \href {https://doi.org/10.3389/fphys.2012.00307}  {  { Front. Physiol. }\textbf  { 3}, {307} (2012)}}. 
    
   \bibitem {tiago2020}  {  {T.~L. Ribeiro},  { S. Yu},  { D.~A. Martin},  { D. Winkowski},  { P. Kanold},  { D.~R. Chialvo}, and  { D. Plenz}, \href {https://doi.org/10.1038/s41598-020-69154-0}  {  { bioRxiv}, {12145} (2020)}}. 
   
 \bibitem{Camargo} S. Camargo, D. A. Martin, E. J. A Trejo, A. de Florian, M. A. Nowak, S.A. Cannas, T.S. Grigera and D. R. Chialvo, 
\href {https://doi.org/10.48550/arXiv.2206.07797} {arXiv:2206.07797 (2022).}
   
 \bibitem{Emiliani2015} V. Emiliani, A. E. Cohen, K. Deisseroth and  M. Häusser, \href{https://doi.org/10.1523/JNEUROSCI.2916-15.2015 }{{J. Neurosci.} \textbf{35}, 13917--13926 (2015)}.
  
 
 \bibitem {BoxScaling}  {  {D.~A. Martin},  { T.~L. Ribeiro},  { S.~A. Cannas},  { T.~S. Grigera},  { D. Plenz} and  { D.~R. Chialvo},\href {https://doi.org/10.1038/s41598-021-95595-2}  {  { Sci. Rep.} \textbf  { 11}, {15937} (2021)}}. 

 \bibitem {Zarepour}    {M. Zarepour},  { J.~I. Perotti},  { O.~V. Billoni},  { D.~R. Chialvo} and  { S.~A. Cannas}, \href {https://doi.org/10.1103/PhysRevE.100.052138}  {  { Phys. Rev. E} \textbf  { 100}, {052138} (2019)}. 
 
\bibitem {Greenberg}    {J.~M. Greenberg} and  { S. Hastings}, \href {https://doi.org/10.1137/0134040}  {  SIAM J Appl Math.\textbf  { 34}, {515} (1978)}. 

   \bibitem{Codes} {The computer codes to generate numerical simulation results and  data analysis can be found} \href{https://github.com/DanielAlejandroMartin/Kappa_C}{\bf Here:}
   \verb|https://github.com/DanielAlejandroMartin/Kappa_C|.

   
\bibitem{Zapperi} S. Zapperi, K. B. Lauritsen, and H. E. Stanley \href{10.1103/PhysRevLett.75.4071} {
Phys. Rev. Lett. \textbf{75}, 4071 (1995)}. 
  
  \bibitem {PowerLaw} A. Deluca  and  A. Corral, \href{https://doi.org/10.2478/s11600-013-0154-9}{ Acta Geophys. \textbf  {61}, 1351–1394 (2013)}. 
 
 
 \bibitem {KappaShew}    W.~L. Shew,   H. Yang,   T. Petermann,  { R. Roy} and  { D. Plenz}, \href {https://doi.org/10.1523/JNEUROSCI.3864-09.2009}  {   J. Neurosci.\textbf { 29}, 15595 (2009)}.
 
   \bibitem{cavagna2010} {A. Cavagna,  A. Cimarelli, I. Giardinaa, G. Parisi, R. Santagati, F. Stefanini, and M. Viale, \href{https://doi.org/10.1073/pnas.1005766107}{ PNAS \textbf{107}  11865–11870  (2010)}}.



 \bibitem {tang}  {  {Q.-Y. Tang},  { Y.-Y. Zhang},  { J. Wang},  { W. Wang} and  { D.~R. Chialvo},\href {https://doi.org/10.1103/PhysRevLett.118.088102}  {  { Phys. Rev. Lett.} \textbf  { 118}, {088102} (2017)}}. 

 \bibitem {tang2}   {Q.-Y. Tang} and  { K. Kaneko}, \href {https://doi.org/10.1371/journal.pcbi.1007670}  {  { PLOS Computational Biology} \textbf  { 16}, {1} (2020)}. 

 \bibitem {cavagna2014}  {  {A. Attanasi},  { A. Cavagna},  { L. Del~Castello},  { I. Giardina},  { S. Melillo},  { L. Parisi},  { O. Pohl},  { B. Rossaro},  { E. Shen},  { E. Silvestri}, and  { M. Viale} \href {https://doi.org/10.1103/PhysRevLett.113.238102}  {  { Phys. Rev. Lett.} \textbf  { 113}, {238102} (2014)}.} 

 \bibitem {flocks}    A. Cavagna,   I. Giardina, and   T.~S. Grigera, \href {https://doi.org/https://doi.org/10.1016/j.physrep.2017.11.003}  {   Phys. Rep. \textbf{ 728}, 1 (2018)}.  
 
 
 \bibitem{grigera} T. S. Grigera, \href {https://doi.org/10.1088/2632-072X/ac2b06} {J. Phys. Complex. \textbf{2}, 045016 (2021)}.



\bibitem {MarianiCorrLength}  B. Mariani,  G. Nicoletti, M. Bisio, M. Maschietto, S. Vassanelli and S. Suweis, \href{https://arxiv.org/abs/2105.05070}{arXiv:2105.05070 (2021)}.   

\bibitem{Comment4750} {To compute $C_W$, in Fig. \ref{Fig3}, and related results in Fig. \ref{Fig4}, we take information for  1/20 of the 95 000 frames used for computation. }

    \bibitem{CommentExp} {Alternatively, we propose $r_0(W_i)=a_i (W_i-W_{min})+r_0(W_{min})$ for experimental data, where $W_{min}$ is the smallest used window.}

 \bibitem {ChenSethna}    Y.-J. Chen,   S. Papanikolaou,   J.~P. Sethna,   S. Zapperi, and   G. Durin \href {https://doi.org/10.1103/PhysRevE.84.061103}  {Phys. Rev. E \textbf  { 84}, {061103} (2011)}. 
 
 
 \bibitem{Control} {D. R. Chialvo}, {S. A. Cannas}, {T. S. Grigera},  {D. A. Martin}, and {D. Plenz} \href{https://doi.org/10.1038/s41598-020-69154-0} 
 {Sci. Rep. \textbf{10} (1), {1-7} (2020)}.
 

 \bibitem{Allen} Allen Institute MindScope Program (2016). Allen Brain Observatory -- 2-photon Visual Coding, experimental dataset i.d. 502368172 from container 511510670, available from brain-map.org/explore/circuits.  Data recorded  (at 30Hz for 114099  time frames) from the VISp area of a conscious mice, corresponding to the inferred spike probabilities of 295 neurons inside a field of view of 400$ \times$400 $\mu m$ and 175 $\mu m$ depth. 
Primary publication: de Vries, S. E. J., Lecoq, J. A., Buice, M. A., et al.  \href{https://doi.org/10.1038/s41593-019-0550-9}{{\em Nat. Neurosci}, \textbf{23}, 138-151 (2020).} 

\bibitem{RatStim} {W. Shew, W. Clawson, J. Pobst, Y. Karimipanah, N. Wright,  and R. Wessel, \href{https://www.nature.com/articles/nphys3370}{{\em Nat. Phys.} \textbf{11}, 659-663 (2015).} }

\bibitem{TurtleStim}  {Z. Ma, G. Turrigiano, R. Wessel,  and K. Hengen,  \href{https://www.sciencedirect.com/science/article/pii/S0896627319307378}{{\em Neuron}. \textbf{104}, 655-664 (2019).}}




\bibitem{Distance} {R. Perin, T. Berger and H. Markram, \href{https://www.pnas.org/doi/abs/10.1073/pnas.1016051108}{{\em PNAS} \textbf{108}, 5419-5424 (2011).}} 
 
 \bibitem {Behtash}  A. Rupasinghe, N. Francis, J. Liu, Z. Bowen, P. O. Kanold, and  B. Babadi, \href{https://doi.org/10.7554/eLife.68046}{eLife \textbf{10}:e68046 (2021)}.   

  \bibitem{Behtash2} M. Shoutik and B. Babadi  \href{https://proceedings.neurips.cc/paper/2021/file/2172fde49301047270b2897085e4319d-Paper.pdf}  
 {in \emph{Advances in Neural Information Processing} \textbf{34} {4120--4133} (2021)}.

  
\end{thebibliography}
\end{document}